\documentclass[aps,twocolumn,
amsmath,prd,superscriptaddress,notitlepage
]{revtex4-1}

\usepackage[utf8]{inputenc}
\usepackage[T1]{fontenc}
\usepackage{dcolumn}
\usepackage{bm}
\usepackage{amsmath}
\usepackage{graphicx} 
\usepackage{tabularx}
\usepackage{multirow}
\usepackage{tikz}
\usepackage{rotating} 
\usepackage{makecell}
\usepackage{amssymb}
\usepackage[normalem]{ulem}
\usepackage{hyperref}
\hypersetup{colorlinks=true, linkcolor=blue, citecolor=blue, urlcolor=blue}
\usepackage{lipsum}
\usepackage{textgreek}
\usepackage{chemformula}
\usepackage{float}
\usepackage{subfiles}
\usepackage{hyperref}

\begin{document}
\title{Stabilization of a non-superconducting, orthorhombic phase by over-hydrogenating LaFeSiH}

\author{M. F. Hansen}
\email{mfh49@cam.ac.uk}
\affiliation{CNRS, Universit\'e Grenoble Alpes, Institut N\'eel, F-38000 Grenoble, France}
\affiliation{Cavendish Laboratory, University of Cambridge, JJ Thomson Avenue, Cambridge, CB3 0HE, United Kingdom}

\author{C. Lepoittevin}
\affiliation{CNRS, Universit\'e Grenoble Alpes, Institut N\'eel, F-38000 Grenoble, France}

\author{J.-B. Vaney}
\affiliation{CNRS, Universit\'e Bordeaux, ICMCB, UPR 9048, F-33600 Pessac, France}

\author{P. Boullay}
\affiliation{Normandie Université, ENSICAEN, UNICAEN, CNRS, CRISMAT, 14050 Caen, France}

\author{V.~Nassif}
\affiliation{CNRS, Universit\'e Grenoble Alpes, Institut N\'eel, F-38000 Grenoble, France}
\affiliation{Institut Laue-Langevin, 71 Avenue des Martyrs, 38000 Grenoble cedex 9, France}

\author{A. Sulpice}
\affiliation{CNRS, Universit\'e Grenoble Alpes, Institut N\'eel, F-38000 Grenoble, France}

\author{H. Mayaffre}
\affiliation{Univ. Grenoble Alpes, CNRS, Univ. Toulouse, INSA-T, LNCMI-EMFL, UPR3228, Grenoble, France}

\author{M.-H. Julien}
\affiliation{Univ. Grenoble Alpes, CNRS, Univ. Toulouse, INSA-T, LNCMI-EMFL, UPR3228, Grenoble, France}

\author{S. Tenc\'e}
\affiliation{CNRS, Universit\'e Bordeaux, ICMCB, UPR 9048, F-33600 Pessac, France}

\author{P. Toulemonde}
\email{pierre.toulemonde@neel.cnrs.fr}
\affiliation{CNRS, Universit\'e Grenoble Alpes, Institut N\'eel, F-38000 Grenoble, France}

\date{\today}

\begin{abstract}
Chemical composition provides a powerful route to tune the electronic ground state of iron-based superconductors and other quantum materials, yet access to highly doped phases remains limited. Here we demonstrate that high-pressure thermal decomposition of hydrogen-rich precursors enables over-hydrogenation of LaFeSi. Using anthracene, we synthesize tetragonal superconducting LaFeSiH, including a single hydrogen site, while ammonia borane yields a structurally distorted over-hydrogenated phase, LaFeSiH$_{1+x}$, with an orthorhombic structure. Chemical analysis reveal excess hydrogen (x~$\approx$~0.6), implying a second H site in LaFeSiH$_{1.6}$ whose localization and occupancy are determined by neutron diffraction. In contrast to metallic LaFeSi and superconducting LaFeSiH, orthorhombic LaFeSiH$_{1.6}$ exhibits semiconductor-like behavior. Upon hydrogen release near 100~\textdegree{}C, it transforms into tetragonal superconducting LaFeSiH$_{1+\delta}$ ($\delta \ll 0.6$). These results establish the chemical flexibility of the layered LaFeSiX (X = H, O, F) family and provide access to a high hydrogen-doping regime, creating new opportunities to investigate superconductivity in Fe-based silicides.

\end{abstract}

\maketitle

\section{Introduction} 

The prototypical iron-based superconductor (IBS) LaFeAsO (1111-type) is a part of a large structural family that includes oxides, fluorides, and hydrides, crystallizing in the ZrCuSiAs-type structure. The expansion of the family of iron-based superconductors (IBS) to the iron crystallogenides LaFeSi(H/F/O) \cite{2018_Bernardini,Vaney2022,Hansen2022}, brought with it prospects of increasing the T$_{c}$ of these materials by tuning the electronic structure through e.g. chemical substitution on the \textit{2b} site of the tetragonal lattice, i.e. the site of H/F/O (at the center of the La$_{4}$ tetrahedra). In the 1111 arsenides, high pressure techniques have been used to make a wide range of substitutions, unveiling the effects of doping on the electronic structure \cite{HOSONO2018278,2013_Hosono,2020Hosono}. This suggests that using high pressure, one might be able to tune the electronic properties of FeSi based superconductors by appropriate chemical substitution. Contrary to arsenides, the silicides show a higher degree of chemical diversity. Substitution on the lanthanum (\textit{2c}) site, a well explored substitution scheme in the arsenides, has recently been explored in La$_{1-x}$Ce$_{x}$FeSiH \cite{Sourd_2026}, however, it is the substitution on the arsenic site which sets the silicides apart. Such substitutions span to all end-members, notably with the hydride LaFeSiH, suggesting that a high tunability should be possible in these systems \cite{2018_Bernardini,2023_Hansen}. Although the application of pressure is not necessary to synthesize the end-members of the series LaFeSiO$_{1-x}$H$_{x}$, the full substitution series has not yet been synthesized. However, a different structural perturbation strategy, is to push LaFeSiH to higher hydrogen contents and investigate the impact on the electronic properties in this system. This could allow doping levels higher than those ever seen in any IBS \cite{2020Hosono}.

Here, we study the synthesis of (metastable) hydrides of LaFeSi using high-pressure techniques, in two different temperature regions, by using two different hydrogen sources, to leverage the kinetics barrier at lower temperatures to stabilize high hydrogen contents. We report a new over-hydrogenated phase, LaFeSiH$_{1.6}$, where the four-fold rotational symmetry of LaFeSiH is broken, reducing the symmetry to orthorhombic. Furthermore, LaFeSiH$_{1.6}$ exhibits a semiconductor-like behavior, as opposed to the metallic precursor LaFeSi and tetragonal superconductor t-LaFeSiH (metallic in the normal state). To assess the thermal stability of LaFeSiH$_{1.6}$ we performed high temperature X-ray powder diffraction~(XRD), revealing that LaFeSiH$_{1.6}$ expels the extra hydrogen around 100~\textdegree{}C and reverts to the tetragonal structure, re-assuming its superconducting properties. We have also followed the decomposition of LaFeSiH$_{1.6}$ to t-LaFeSiH using thermogravimetric analysis (TGA), differential thermal analysis (DTA) and mass spectrometry (MS), confirming the scenario that the extra hydrogen leaves the structure upon heating. Combining transmission electron microscopy (TEM) and neutron powder diffraction (NPD) allows us to determine the localization of the two H sites in LaFeSiH$_{1.6}$. Finally, by comparison with t-LaFeSiH which has only a single H site, nuclear magnetic resonance (NMR) measurements confirms the existence of the second extra hydrogen site. 

\section{Methods}
\subsection{Synthesis}
Synthesis was carried out using LaFeSi as a precursor, prepared as described by Bernardini et al. \cite{2018_Bernardini}. The precursor material typically had a purity of $\sim$ 95-97~\% with minor Fe-based impurities being paramagnetic LaFe$_{2}$Si$_{2}$ ($\sim$ 2~wt~\%)~\cite{UMARJI198361,Rogl1989} and magnetic La(Fe$_{1-x}$Si$_{x}$)$_{13}$ (< 1~wt~\%)~\cite{PALSTRA1983290} (see precursor XRD pattern in Suppl. Mat. of Ref.~\cite{2023_Hansen}). Such minor secondary phases remain after the hydrogenation process with the second one being hydrogenated into magnetic La(Fe$_{1-x}$Si$_{x}$)$_{13}$H$_{y}$ (undetectable in our laboratory XRD) which orders below a Curie temperature above room temperature (up to $\sim$~340~K)~\cite{PHEJAR201695}. These tiny impurities do not affect the conclusions made for the hydrogenated main phase LaFeSiH$_{1+x}$ studied here. However, the presence of a minute amount of La(Fe$_{1-x}$Si$_{x}$)$_{13}$H$_{y}$ gives a magnetic background which complicates the direct observation of superconductivity in magnetization (See Supplemental Material.~B \cite{SuppInf_B}).
To make our high pressure crucible assembly, dried NaCl was compacted in a 9~mm die, applying 8~tons, to form pellets with a height of around 1.5~cm and 0.5~cm. The tall pellets were machined into crucibles with an inner diameter of 6~mm. All pellets were subsequently dried by placing them in a furnace at 200~\textdegree C for 24~h to ensure that water was not present in the crucible. The finished crucibles were then transferred to a glove box for dry storage under Argon. 
When preparing the synthesis, pellets of LaFeSi and Hydrogen source material (Anthracene or Ammonia Borane (AB)) were pressed in two different pellets with mass ratios to have excess hydrogen corresponding to H/LaFeSi molar ratio of approximately 20/1, with a diameter matching the bore of the NaCl crucibles. A NaCl crucible was then transferred into a Au capsule and both pellets were inserted in the NaCl crucible separated physically by a thin hexagonal BN cylindrical spacer which is known to prevent carbon diffusion~\cite{DOSSANTOS1999L1}, in order to avoid any reaction of LaFeSi with C or B, from the decomposed hydrogen source, Anthracene and AB respectively. The NaCl crucible was then closed with a NaCl lid, where-after the Au capsule was sealed with an opposing Au capsule piece. All of the above steps were performed under Argon. The capsule was transferred to a CONAC-28 type press and pressurized to 0.5~-~1~GPa for Anthracene and 1~GPa for AB, ensuring a tight seal of the assembly. The sample was then heated to 600~-~700~\textdegree{}C for Anthracene and 400~\textdegree{}C for AB and subsequently quenched to room temperature. The pressure was slowly released and the capsule was recovered. A LaFeSiH$_{1+x}$ sample was prepared with AB for a dedicated neutron powder diffraction experiment using similar conditions (1-1.5~GPa and 400~\textdegree{}C) and 20/1 molar ratio in a CONAC-40 type press, which allows to treat larger amount of reactants. After checking by X-ray diffraction (XRD), two batches were mixed yielding a total amount of $\sim$ 450~mg for this NPD sample. Using AB allows for hydrogen release at lower temperatures than when using Anthracene due to their different decomposition temperatures. While AB decomposes at lower temperature, the decomposition is complex in comparison to that of Anthracene which decomposes directly to graphite at $\simeq{}$~550~\textdegree{}C in the 1-10~GPa range \cite{CHANYSHEV2017,Frueh2011}.
When using Anthracene the ends of the capsules had a convex shape  and audible gas escape occurred upon puncturing the Au capsule. When using AB the convex shape as well as the gas escape were not observed, however, a strong odor was present upon puncturing the capsule, hence, the capsule was allowed to degas in a fume hood.

\subsection{X-ray diffraction}
XRD was performed using a Bruker D8 Endeavor diffractometer in reflection geometry and using a Bruker D5000 transmission diffractometer both equipped with a Cu-source ($\lambda$ = 1.54~\AA) in the the 10~-~90\textdegree~$2\theta$~range.

\subsection{Electron diffraction}
For unit cell and space group determination by TEM studies, a specimen was prepared by crushing a small amount of powder in an agate mortar containing ethanol, and a drop of the homogeneous suspension was deposited onto a holey carbon membrane supported by a copper grid. Zone axis electron diffraction patterns were recorded on a Philips CM300ST microscope (LaB6, 300 kV) equipped with a 4K TVIPS CMOS camera.

Precession-assisted 3D electron diffraction (3D ED) \cite{2019_Gemmi}, was performed with a JEOL F200 transmission electron microscope (200 kV - cold FEG) equipped with a Nanomegas DigiStar precession module and a bottom-mounted ASI Cheetah M3 hybrid-pixel detector. 3D ED aims a three-dimensional reconstruction of the reciprocal space by tilting sequentially the sample and recording at each step a precession electron diffraction patterns. 3D ED data sets of non-oriented patterns were recorded on several crystals. For data collections the precession angle was set to 1.2\textdegree{} with a 1\textdegree{} goniometer tilt step. 3D ED data sets were analysed and diffracted intensities integrated using the program PETS 2.0 \cite{2019_Palatinus}. The structure solution, achieved via ab initio phasing by charge-flipping, and subsequent structure refinements, incorporating both dynamical diffraction effects and precession motion, were carried out using JANA2020 \cite{2023_Petricek}.

\subsection{Neutron diffraction}
Neutron powder diffraction was measured at room temperature using the D1B instrument of Institut Laue Langevin (Grenoble, France). 
The powder was placed in a Vanadium sample holder and inserted in the beam. The NPD pattern was collected for two hours using a monochromatic neutron beam at $\lambda$ = 1.287 \AA{} with $2\theta$ in the 5~-~128\textdegree~range.

\subsection{Thermogravimetric Analysis, Differential Thermal Analysis, and Mass Spectroscopy}
For the TGA/DTA/MS (Thermograviometry / Differential thermal analysis / Mass spectroscopy) measurements, a 70 $\mu$L alumina crucible was filled with 181.1~mg of orthorhombic o-LaFeSiH$_{1+x}$ and placed in a symmetrical Setaram TAG 16 TGA/DTA system coupled with a Hiden QGA mass spectrometer. The sample and compensation chambers were kept in the same constant gas flow of Ar to avoid any mass drift due the Archimedes force of the gas flow on the sample. The temperature was increased using a ramp of 10 \textdegree{}C/min to a maximum temperature of 500 \textdegree{}C. During the experiment, the relative abundance of a large range (m/z = 1-50) in the exhausted gas were measured by the MS system.

\subsection{Electrical resistivity}
Resistance as a function of temperature was measured in the four probe geometry down to 4.2 K with an excitation current of 100 $\mu$A using a custom build He$^{4}$ cryostat. Current was provided using a Keithley 6220 precision current source and the resulting voltage was measured using a Keithley 2182A Nano-voltmeter.

\subsection{NMR measurements}
$^{29}$Si nuclear magnetic resonance (NMR) measurements were performed in a 15~T superconducting coil, using a home-built heterodyne spectrometer. The field value was calibrated using metallic Cu from the NMR pick-up coil. Spectra were obtained by adding appropriately-spaced Fourier transforms of the spin-echo signal. For the desorbed t-LaFeSiH$_{1+\delta}$ sample, the spin-lattice relaxation time $T_1$ was measured as a function of temperature by the saturation-recovery method and the recoveries were fit to a single exponential function (as expected for magnetic relaxation of a nuclear spin 1/2) with an ad-hoc stretching exponent $\beta\simeq 0.8-0.9$ accounting for the distribution of $T_1$ values.

\section{Results and discussion}
\subsection{Formation and crystal structure of orthorhombic LaFeSiH$_{1+x}$}

Samples were synthesized using LaFeSi as a precursor and introducing hydrogen at high pressure using thermal decomposition of a hydrogen source. The crystal structure of samples made using both Anthracene and AB as hydrogen sources were probed using XRD.

\begin{figure}[!b]
    \centering
    \includegraphics[width = \linewidth]{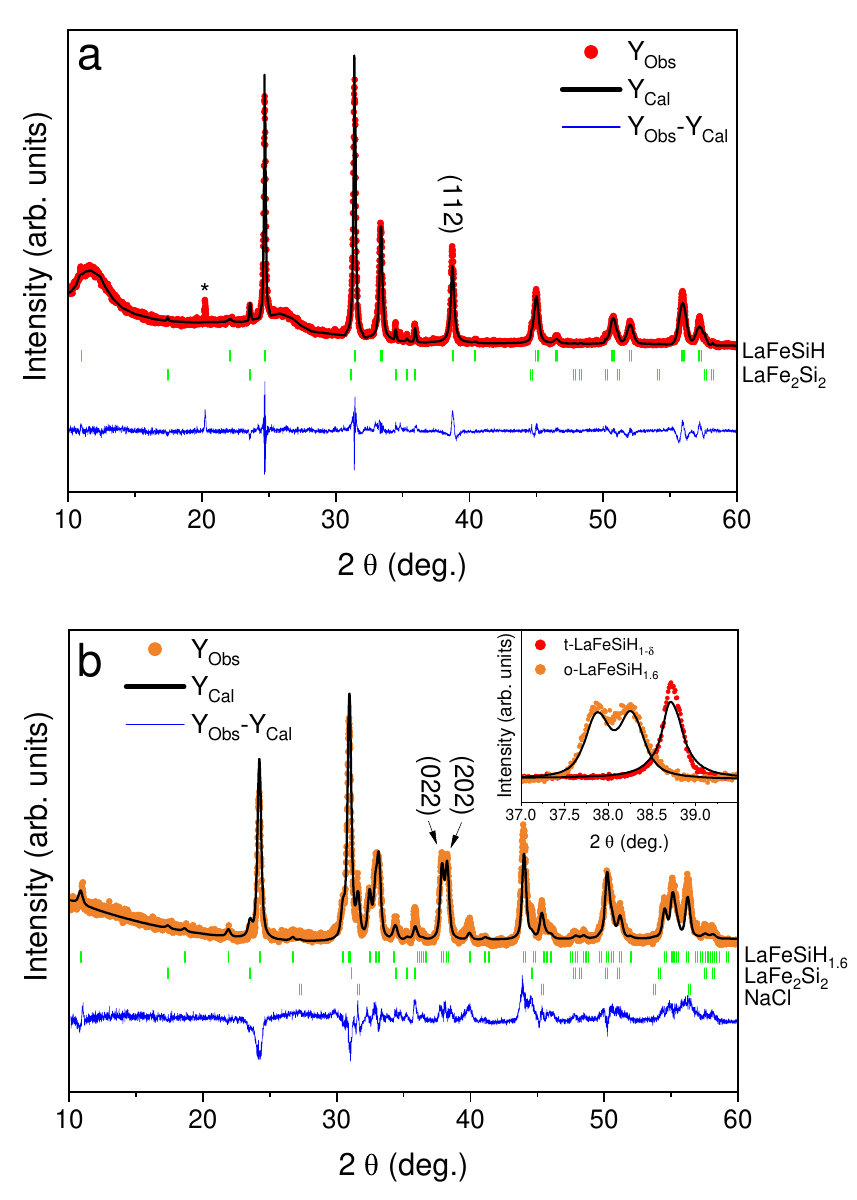}
    \caption{X-ray diffraction patterns, in red and orange respectively, collected for samples synthesized at high pressures - high temperature (HP-HT) using, (a), Anthracene as a hydrogen source, and, (b), using AB as a hydrogen source. A Rietveld fit of the data is shown (in black), based respectively on the t-LaFeSiH structure from Bernardini et al. \cite{2018_Bernardini} in (a), and on an orthorhombic structural model in the space group \textit{Pmab} in (b), for the main phase. The peak marked with a asterisk in the (a) is a spurious peak. The inset in (b) shows an enlarged view comparing both patterns highlighting the induced orthorhombic distortion when AB is used at HP-HT.}
    \label{fig:XRD_synthesis}
\end{figure}

 In figure \ref{fig:XRD_synthesis}.a the XRD pattern of a sample made using Anthracene has been plotted along with a corresponding Rietveld fit. Here, the LaFeSiH atomic positions published by Bernardini et al. have been assumed \cite{2018_Bernardini} and are not refined. From the fit, we obtain the tetragonal unit cell parameters \textit{a} = \textit{b} = 4.027(1)~\AA{} and \textit{c} = 8.038(2)~\AA{}, which is consistent with the previously reported unit cell parameters \cite{2018_Bernardini,2023_Hansen,2024_Hansen}. We observe superconducting behavior with T$_{c~onset}$ from 5~K to 8.5~K (see Supplemental information.~A \cite{SuppInf_A}), in both resistivity and magnetization measurements, i.e. similar to the one for tetragonal t-LaFeSiH prepared by conventional hydrogenation \cite{2018_Bernardini,2023_Hansen,2024_Hansen}. Samples synthesized using Anthracene will not be discussed further as the resulting phase is well documented.

In figure \ref{fig:XRD_synthesis}.b we show the XRD pattern obtained from the sample made at lower temperatures using AB as the hydrogen source. The pattern bears some qualitative similarities with the pattern in figure \ref{fig:XRD_synthesis}.a, owning to similar d-spacings in the two compounds. There is however a clear splitting of several peaks, most clearly the splitting of the (112) t-LaFeSiH reflection into two peaks, as shown in the inset of figure \ref{fig:XRD_synthesis}.b. Upon examining the pattern, it is clear that the split peaks are all (\textit{hhl}), and that the system has undergone an orthorhombic distortion. The broken symmetry is likely due to a higher hydrogen content (x). This would require the existence of a second hydrogen site, in addition to the (\textit{2b} site) found in stoichiometric t-LaFeSiH. We denote this new phase o-LaFeSiH$_{1+x}$.

\begin{figure}[!b]
    \centering
    \includegraphics[width = \linewidth]{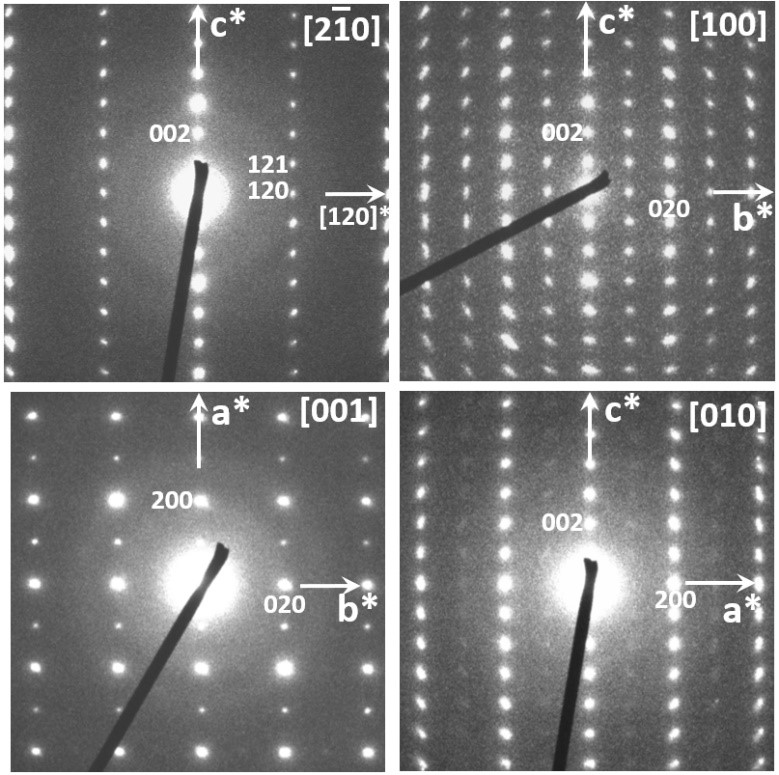}
    \caption{Zone axis ED patterns of o-LaFeSiH$_{1+x}$ indexed in the orthorhombic space groups \textit{Pmab} or \textit{P}2$_{1}$\textit{ab}.}
    \label{fig:TEM_SG}
\end{figure}

To further investigate this distortion we performed an electron diffraction study on grains of o-LaFeSiH$_{1+x}$. The analysis of zone axis electron diffraction patterns reveal an orthorhombic unit cell defined by the lattice parameters \textit{a} $\approx$ 5.8 Å, \textit{b} $\approx$ 5.9 Å and \textit{c} $\approx$ 8.1 Å. The indexation of the reflections evidenced an absence of systematic extinction conditions on \textit{hkl} and 0\textit{kl} reflections, as well as systematic extinctions on \textit{h}0\textit{l}~: \textit{h} = 2n and \textit{hk}0~: \textit{k} = 2n (Figure \ref{fig:TEM_SG}). Thus the possible resulting space groups are \textit{Pmab} (57) or \textit{P}2$_{1}$\textit{ab} (29). It has to be noted that \textit{h}00 and 0\textit{k}0 reflections with \textit{h} and \textit{k} odd should be extinguished and their appearance on [001] and [100] electron diffraction patterns is due to multiple diffraction.

\begin{figure}
    \centering
    \includegraphics[width = \linewidth]{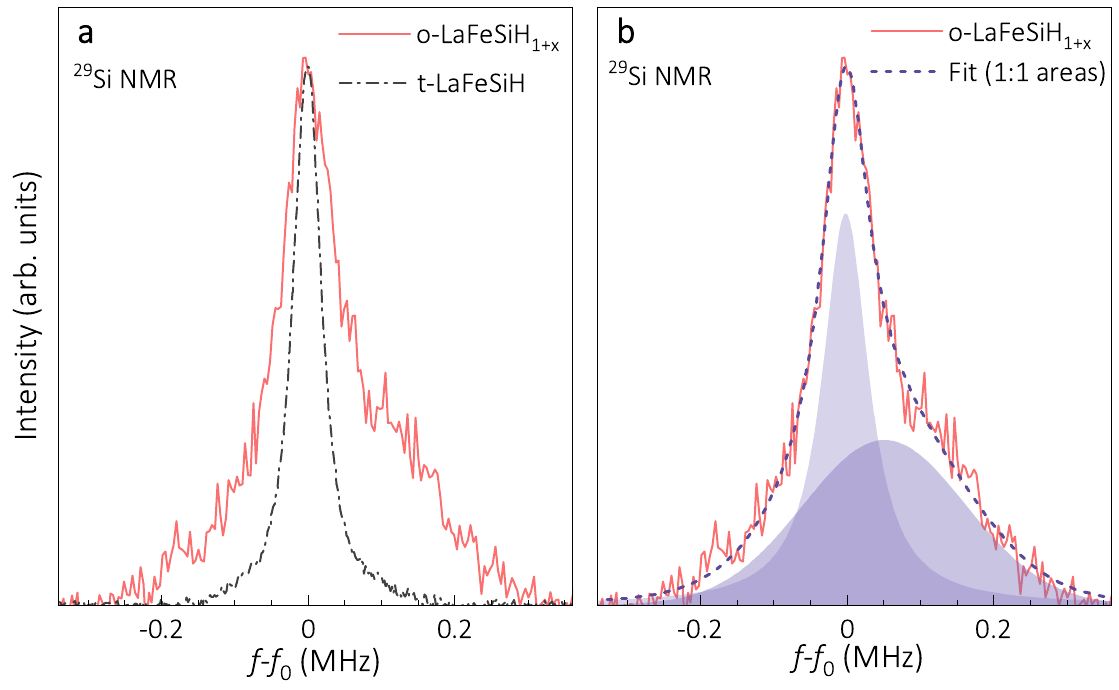}
    \caption{(a) $^{29}$Si NMR spectra of orthorhombic LaFeSiH$_{1+x}$ (solid line, $T=4.9$ K, $f_0=126.75$ MHz, this work) and tetragonal LaFeSiH (dash–dotted line, $T=2.8$ K, $f_0=126.79$ MHz, from Ref.~\cite{2024_Hansen}), both measured in a magnetic field of 15 T. The o-LaFeSiH$_{1+x}$ spectrum exhibits a pronounced high-frequency shoulder. (b) Fit of the o-LaFeSiH$_{1+x}$ spectrum to the sum of two Voigt profiles with constrained equal areas. Fits using Lorentzian profiles yield comparable agreement.}
    \label{fig:NMR1}
\end{figure}

Our NMR results support bulk chemical modification in o-LaFeSiH$_{1+x}$. The $^{29}$Si NMR spectrum (Fig.~\ref{fig:NMR1}) exhibits a high-frequency shoulder absent in t-LaFeSiH. The spectrum of o-LaFeSiH$_{1+x}$ can be reproduced using two peaks (Fig.~\ref{fig:NMR1}). The substantial spectral weight of this feature, together with its disappearance upon thermal decomposition (see below), strongly argues against a spurious secondary phase.

The same shoulder is observed at 5~K (Fig.~\ref{fig:NMR1}) and 50 K, indicating that it is unlikely to originate from a temperature-dependent electronic phenomenon. In a random powder, magnetic order would broaden the NMR line symmetrically rather than produce a distinct shoulder—except in the unlikely scenario of a large ferromagnetic fraction coexisting with a paramagnetic one. Likewise, while an orthorhombic distortion can split the NMR line in a single crystal (see Ref.~\cite{Zhou2020} for an Fe-chalcogenide example), it would not generate such an asymmetric line shape in a powder sample.

The most natural interpretation of the line shape is the presence of two Si sites with distinct local environments in o-LaFeSiH$_{1+x}$. However, any specific site assignment remains premature. Owing to the limited signal-to-noise ratio (due to the small sample mass) and the small peak separation relative to the intrinsic line width, fits of the $^{29}$Si spectrum to two components suffer from parameter degeneracy. This leads to underestimated uncertainties and strong sensitivity to initialization conditions. While Fig.~\ref{fig:NMR1} displays a representative fit constrained to equal areas, we show in Supplemental information.~C \cite{SuppInf_C} that statistically equivalent fits ($\chi^2 \approx 0.99$) can be obtained with unconstrained peak areas that differ substantially, and even invert (area ratios ranging from 1:3 to 3.7:1, see Supplemental information.~C \cite{SuppInf_C}). The inability to determine a reliable relative area ratio therefore precludes a quantitative interpretation of the spectra.

Additional $^1$H or $^{139}$La NMR measurements could help provide quantitative insight. Such experiments lie beyond the scope of the present work, whose primary aim is to establish the existence of a new over-hydrogenated member of the LaFeSiH family. Nevertheless, the NMR data are qualitatively consistent with the incorporation of additional hydrogen into the structure.

\subsection{Electrical resistance of o-LaFeSiH$_{1+x}$}

\begin{figure}[!t]
    \centering
    \includegraphics[width = \linewidth]{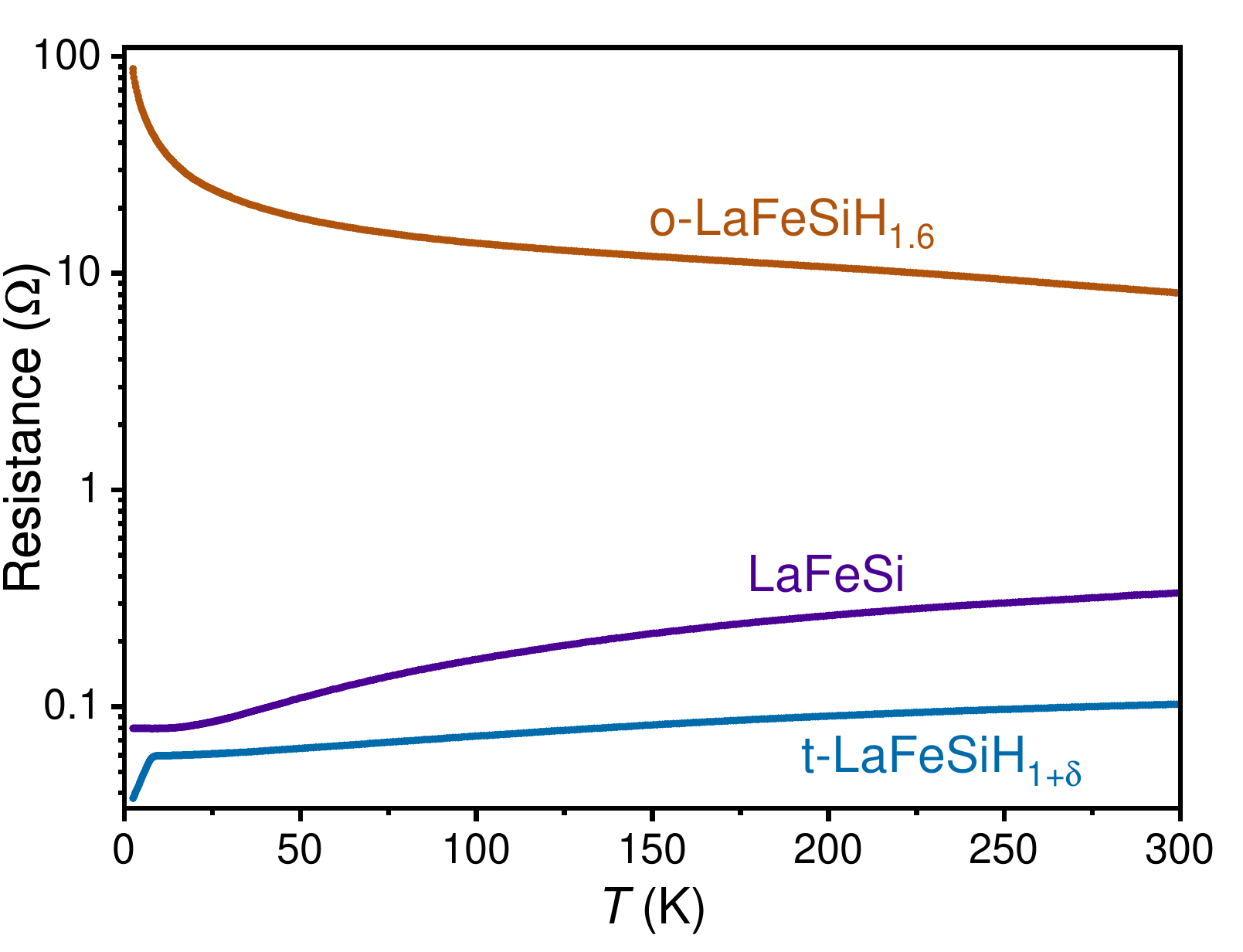}
    \caption{The electrical resistance of LaFeSi (precursor), o-LaFeSiH$_{1+x}$ (over-hydrogenated, orthorhombic) and t-LaFeSiH$_{1+\delta}$ (desorbed), as a function of temperature. The magnetic-field dependence of the data for the three samples, especially supporting the superconducting origin of the low temperature drop for LaFeSiH$_{1+\delta}$, is shown in SI.~D~\cite{SuppInf_D}.}
    \label{fig:resistance}
\end{figure}

(Over-)hydrogenation dramatically alters the electronic properties: the electrical resistance of o-LaFeSiH$_{1+x}$ exhibits semiconducting-like behavior over the entire temperature range (Fig.~\ref{fig:resistance}), in sharp contrast to both the precursor LaFeSi, a nonsuperconducting metal down to at least 2~K, and t-LaFeSiH, which shows metallic conductivity and becomes superconducting below 10~K. Magnetic field effects are shown in Supplemental information.~D \cite{SuppInf_D}.

These results establish that LaFeSiH$_{1+x}$ and LaFeSiH are distinct electronic phases.

\subsection{Insights from thermally decomposing o-LaFeSiH$_{1+x}$}

To better understand the orthorhombic phase and clarify its chemical nature, we investigated thermally decomposed o-LaFeSiH$_{1+x}$. 

\begin{figure*}[!ht]
    \centering
    \includegraphics[width = \linewidth]{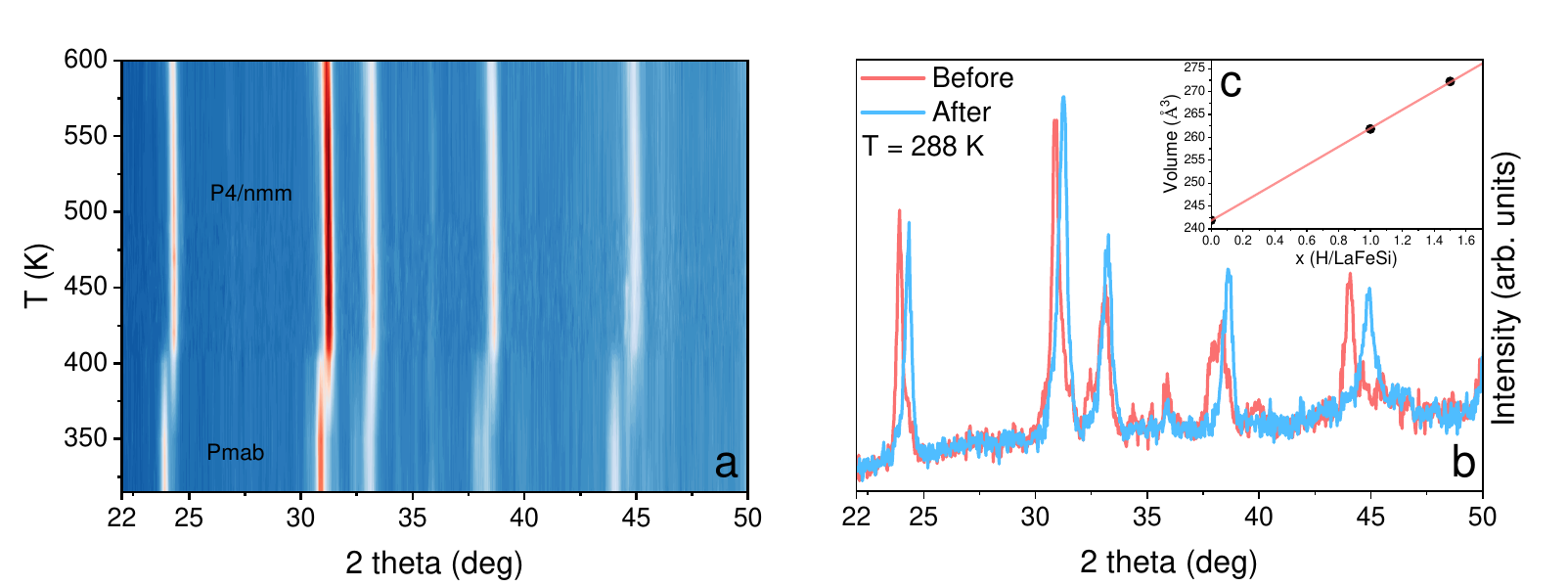}
    \caption{(a): The X-ray diffraction patterns collected at different temperatures during the thermal decomposition of o-LaFeSiH$_{1+x}$. (b): X-ray diffraction patterns collected before and after the heating cycle, with the recovered phase corresponding to t-LaFeSiH. (c): The volume of the unit cell for different amounts of hydrogen in the LaFeSi lattice.}
    \label{fig:HT-XRD}
\end{figure*}

We first performed an in situ XRD experiment. Around 100\textdegree{}C, the XRD pattern (Fig.~\ref{fig:HT-XRD}a and ~\ref{fig:HT-XRD}b) reveals a transition from an orthorhombic structure to a tetragonal phase with \textit{P}4\textit{/nmm} symmetry and a unit cell consistent with t-LaFeSiH. After cooling back to room temperature, the recovered sample does not revert to the orthorhombic phase but retains a diffraction pattern similar to that of t-LaFeSiH. The transition is therefore irreversible, indicating that the orthorhombic and tetragonal phases are chemically distinct.

\begin{figure}[!b]
    \centering
    \includegraphics[width = \linewidth]{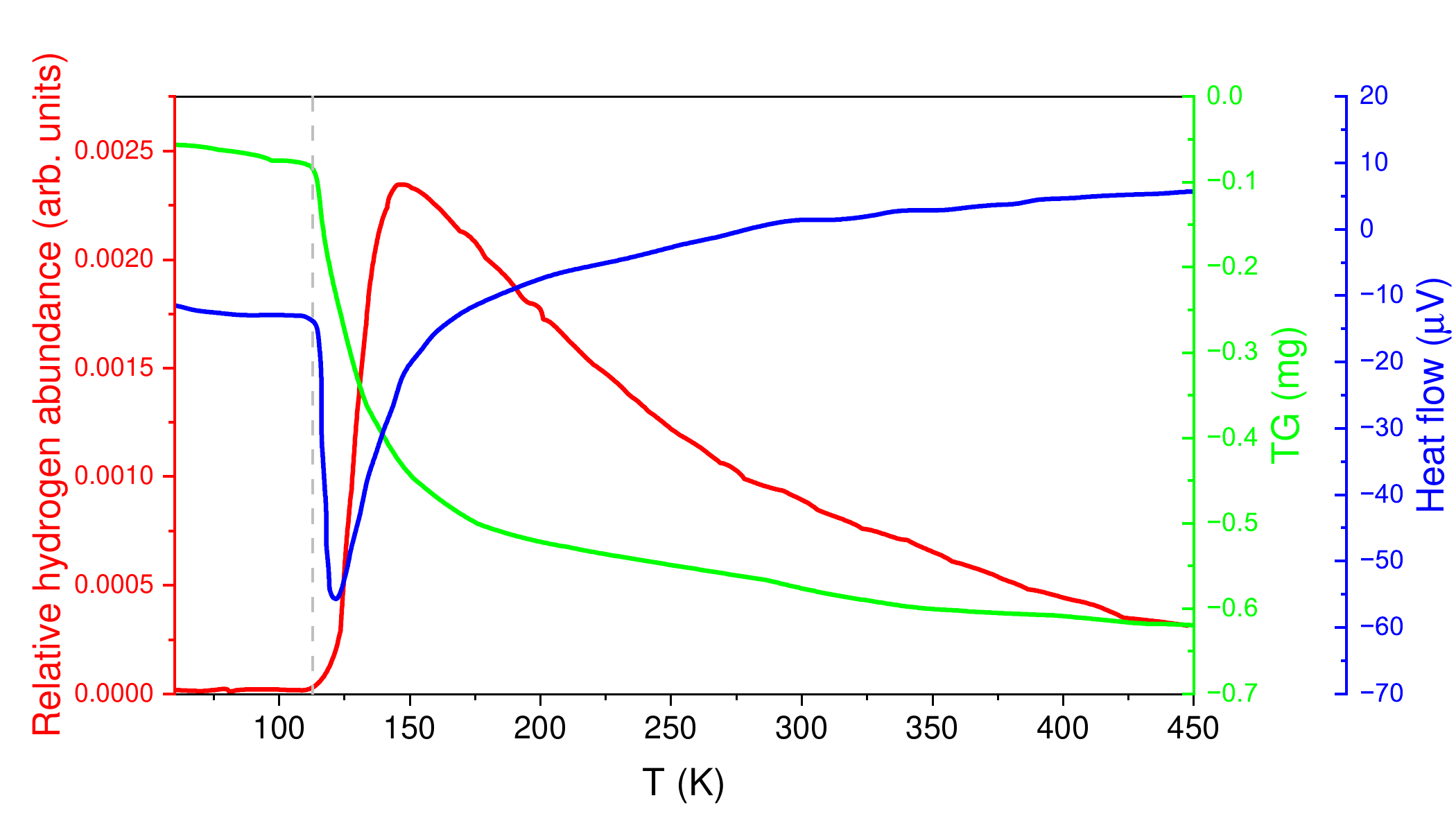}
    \caption{Thermogravimetric analysis, differential thermal analysis and mass spectroscopy measured as a function of temperature during the thermal decomposition of o-LaFeSiH$_{1+x}$}
    \label{fig:TGA}
\end{figure}

To investigate the transition near 100\textdegree{}C, we performed TGA/DTA/MS measurements. Around 110\textdegree{}C, pronounced anomalies are observed in all signals (Fig.~\ref{fig:TGA}). First, the mass decreases by about 0.50(1)~mg (i.e. 0.27~\%). Concomitantly, the relative hydrogen signal in the exhaust gas increases, indicating that the lost mass corresponds to hydrogen released from the sample. No signal was detected in other channels corresponding, for example, to oxygen-bearing species, ruling out moisture loss as the origin of the mass change. The dehydrogenation scenario is further supported by the peak in the heat-flow signal (with an opposite sign compared to the one for hydrogenation of LaFeSi into LaFeSiH \cite{2023_Hansen}), since the desorption of hydrogen could be expected to be an endothermic reaction owing to the enthalpy change, as for many other hydrides \cite{ZHENG_2025}.

Next, we quantify the amount of hydrogen released from the system. If this mass loss is attributed to hydrogen desorption and compared to the initial sample mass, it corresponds to a loss of about 0.61 hydrogen atoms per LaFeSiH formula unit, suggesting that the orthorhombic phase has an approximate composition o-LaFeSiH$_{1.6}$.
Next, we quantify the amount of hydrogen released from the system. If this mass loss is attributed to hydrogen desorption and compared to the initial sample mass, it corresponds to a loss of about 0.61 hydrogen atoms per LaFeSiH formula unit, suggesting that the orthorhombic phase has an approximate composition o-LaFeSiH$_{1.6}$.

\begin{figure*}[!t]
    \centering
    \includegraphics[width = \linewidth]{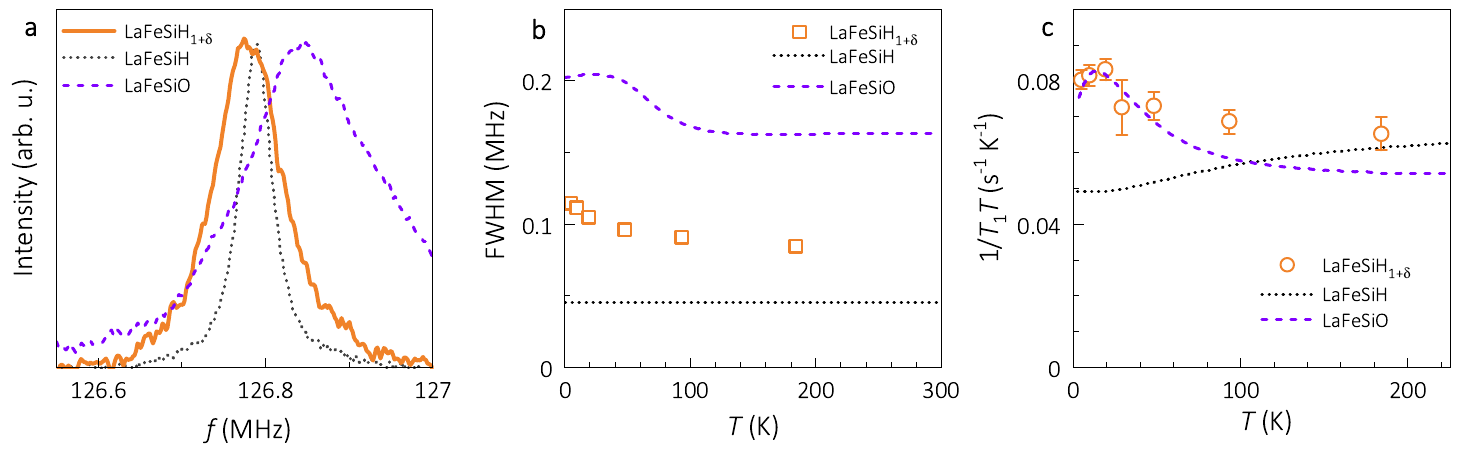}
    \caption{(a) $^{29}$Si NMR spectra of desorbed LaFeSiH$_{1.6}$ (this work), LaFeSiH~\cite{2024_Hansen} and LaFeSiO~\cite{Hansen2022}. (b) Full-width-at-half-maximum of the $^{29}$Si spectra as a function of temperature for the three same samples~\cite{Hansen2022,2024_Hansen}. (c) $1/(T_1T)$ vs. $T$ for the same samples~\cite{Hansen2022,2024_Hansen}.}
    \label{fig:NMR2}
\end{figure*}

The sample measured in Fig.~\ref{fig:NMR1} was heated to 120\textdegree{}C for 6 hours and subsequently re-examined by NMR. The resulting $^{29}$Si spectrum (Fig.~\ref{fig:NMR2}a) no longer displays the high-frequency shoulder characteristic of o-LaFeSiH$_{1.6}$, consistent with desorption of the excess hydrogen. The single resonance line retains a slight asymmetry, being somewhat broader on the high-frequency side--a feature absent in t-LaFeSiH--which suggests that a small fraction ($\delta$) of excess hydrogen likely remains.

Further support for the conclusion that the recovered sample is not pristine LaFeSiH but rather t-LaFeSiH$_{1+\delta}$, with $\delta \ll 0.6$, comes from the reappearance of superconductivity at $T_c \simeq 8$~K (Fig.~\ref{fig:resistance} and Supplemental information.~D \cite{SuppInf_D}), only about 2~K below that of pristine LaFeSiH. Both slight residual over-hydrogenation and vacancies on the \textit{2b} hydrogen site (see below) may account for the remaining difference with LaFeSiH. 

Notably, both the NMR line width (Fig.~\ref{fig:NMR2}b) and the spin-lattice relaxation time $T_1$ (Fig.~\ref{fig:NMR2}c) exhibit temperature dependences that differ from those observed in LaFeSiH~\cite{2024_Hansen}. At low temperatures, these trends resemble those reported for LaFeSiO~\cite{Hansen2022}, where they were attributed to weak magnetic correlations (both the spectra and the $T_1$ data being inconsistent with magnetic order), that are absent in LaFeSiH. At present, it remains unclear whether the differences between LaFeSiH and desorbed t-LaFeSiH$_{1+\delta}$ arise from variations in carrier concentration, disorder, or a combination of both.

Finally, we measured XRD patterns of the precursor LaFeSi, the orthorhombic compound o-LaFeSiH$_{1.6}$, and the recovered tetragonal sample t-LaFeSiH$_{1+\delta}$ investigated by TGA/DTA/MS. The patterns were analyzed using Le Bail refinements to extract the unit-cell volumes. For the tetragonal phases, a pseudo-orthorhombic cell volume was calculated using $a_T = \sqrt{2}\,a_O$ to enable direct comparison. From this analysis, insertion of one hydrogen atom per LaFeSi formula unit (i.e., going from LaFeSi to t-LaFeSiH) results in a volume increase of 19.94~\AA$^{3}$, consistent with previous reports~\cite{2018_Bernardini,2023_Hansen}. As shown in Fig.~\ref{fig:HT-XRD}c, the volume increase from LaFeSi to o-LaFeSiH$_{1.6}$ amounts to 30.31~\AA$^{3}$, corresponding to 1.52 times the increase observed upon formation of t-LaFeSiH. This provides further support for the o-LaFeSiH$_{1.6}$ composition inferred from TGA.

\subsection{Localizing extra hydrogen in o-LaFeSiH$_{1.6}$}

\begin{table*}[!htbp]
\begin{tabular}{cccccccccc}
\hline
\multicolumn{1}{c}{}&\multicolumn{4}{|c|}{Space Group \textit{Pmab} (57)} & \multicolumn{5}{c}{Space Group \textit{Pmab} (57)}\\
\multicolumn{1}{c}{}&\multicolumn{4}{|c|}{a = 5.744(2) \AA, b = 5.825(3) \AA, c = 8.113(5) \AA} & \multicolumn{5}{c}{a = 5.753(1) \AA, b = 5.854(1) \AA, c = 8.083(2) \AA}
\\ \hline
\multicolumn{1}{c}{}&\multicolumn{4}{|c|}{ R(obs) = 10.91~\%, R(all)~= 12.01~\%}  & \multicolumn{5}{c}{Conventional Rietveld R-factors:}      \\ 
\multicolumn{1}{c}{}&\multicolumn{4}{|c|}{ wR(all)~=~22.37~\%, GoF(obs)~=~3.83}  & \multicolumn{5}{c}{Rp~=~ 25.1~\%, Rwp~=~21.1~\%}      \\ 
\multicolumn{1}{c}{}&\multicolumn{4}{|c|}{ meas. / obs. {[}I\textgreater{}3σ(I){]} reflections~=~1557 / 1260 }  & \multicolumn{5}{c}{R$_{exp}$~=~3.88~\%} \\
\multicolumn{1}{c}{}&\multicolumn{4}{|c|}{ 68 refined parameters }  & \multicolumn{5}{c}{Main phase Bragg R-factor~=~12.2~\%} \\
\multicolumn{1}{c}{}&\multicolumn{4}{|c|}{g$_{max}$ (\AA$^{-1}$)~=~1.7, S$_{g,max}$ (\AA$^{-1}$)~=~0.01}  & \multicolumn{5}{c}{R$_{F}$-factor~=~7.43~\%} \\
\multicolumn{1}{c}{}&\multicolumn{4}{|c|}{R$_{Sgmax}$~=~0.4, steps~=~256}  & \multicolumn{5}{c}{} \\ \hline
\multicolumn{1}{c}{Atom label (Wyckoff)}   & \multicolumn{1}{|c}{x} & \multicolumn{1}{c}{y} & \multicolumn{1}{c}{z}  & \multicolumn{1}{c|}{Uiso/Ueq~(\AA$^2$)}  & \multicolumn{1}{c}{x} & \multicolumn{1}{c}{y} & \multicolumn{1}{c}{z}  & \multicolumn{1}{c}{Biso~(\AA$^2$)} & \multicolumn{1}{c}{occ.}   \\ \hline
\multicolumn{1}{c}{La1 (4d)}  &  \multicolumn{1}{|c}{0.25} & \multicolumn{1}{c}{0.4769(2)} & \multicolumn{1}{c}{0.8218(3)}  & \multicolumn{1}{c|}{0.0084(6)$^{\star}$}   & \multicolumn{1}{c}{0.25}   &  \multicolumn{1}{c}{0.477(3)}   & \multicolumn{1}{c}{0.823(1)}   & \multicolumn{1}{c}{0.9(1)}   & \multicolumn{1}{c}{1}       \\
\multicolumn{1}{c}{Fe1 (4c)}  & \multicolumn{1}{|c}{0}  & \multicolumn{1}{c}{0.25} & \multicolumn{1}{c}{0.4847(5) } & \multicolumn{1}{c|}{0.0048(5)}   & \multicolumn{1}{c}{0}     & \multicolumn{1}{c}{0.25}   & \multicolumn{1}{c}{0.490(2)}      & \multicolumn{1}{c}{0.2(1)}  & \multicolumn{1}{c}{1}   \\
\multicolumn{1}{c}{Si1 (4d)}   & \multicolumn{1}{|c}{0.75}   & \multicolumn{1}{c}{-0.0138(8)}   & \multicolumn{1}{c}{0.3788(7)}  & \multicolumn{1}{c|}{0.0066(9)}  & \multicolumn{1}{c}{0.75}   & \multicolumn{1}{c}{-0.022(7)}     & \multicolumn{1}{c}{0.363(3)}    & \multicolumn{1}{c}{2.6(4)}    & \multicolumn{1}{c}{1}    \\
 \multicolumn{1}{c}{H1 (4c)}   & \multicolumn{1}{|c}{0}         & \multicolumn{1}{c}{0.25} & \multicolumn{1}{c}{0.031(14)}  & \multicolumn{1}{c|}{0.05(2)}              & \multicolumn{1}{c}{0}    & \multicolumn{1}{c}{0.25}   & \multicolumn{1}{c}{0}     & \multicolumn{1}{c}{0.8(3)}  & \multicolumn{1}{c}{0.71(4)}   \\
 \multicolumn{1}{c}{H2 (4b)}  &\multicolumn{1}{|c}{-}  & \multicolumn{1}{c}{-} & \multicolumn{1}{c}{-} &\multicolumn{1}{c|}{-}   & \multicolumn{1}{c}{0.75}   & \multicolumn{1}{c}{0.447(5)}   & \multicolumn{1}{c}{0.831(3)}  & \multicolumn{1}{c}{0.8(3)}  & \multicolumn{1}{c}{0.93(4)}   \\ \hline
\end{tabular}
\caption{Refined structural parameters of the LaFeSiH$_{1.6}$ phase. Left column: refinement based on 3D ED data performed in the \textit{Pmab} space group with composition LaFeSiH (considering only one H site (H1)) with full site occupancies, yielding an orthorhombic structure which is analogous to t-LaFeSiH. $^{\star}$Anisotropic ADPs were refined for La1. Right column: refinement based on NPD data; to account for the extra hydrogen we introduce a second hydrogen site, H2 (localized in the Lanthanum layer), and both H1 and H2 site occupancies are refined.}
\label{table:Pmba_ED_plus_NPD}
\end{table*}

\begin{figure}[!t]
    \centering
    \includegraphics[width=\linewidth]{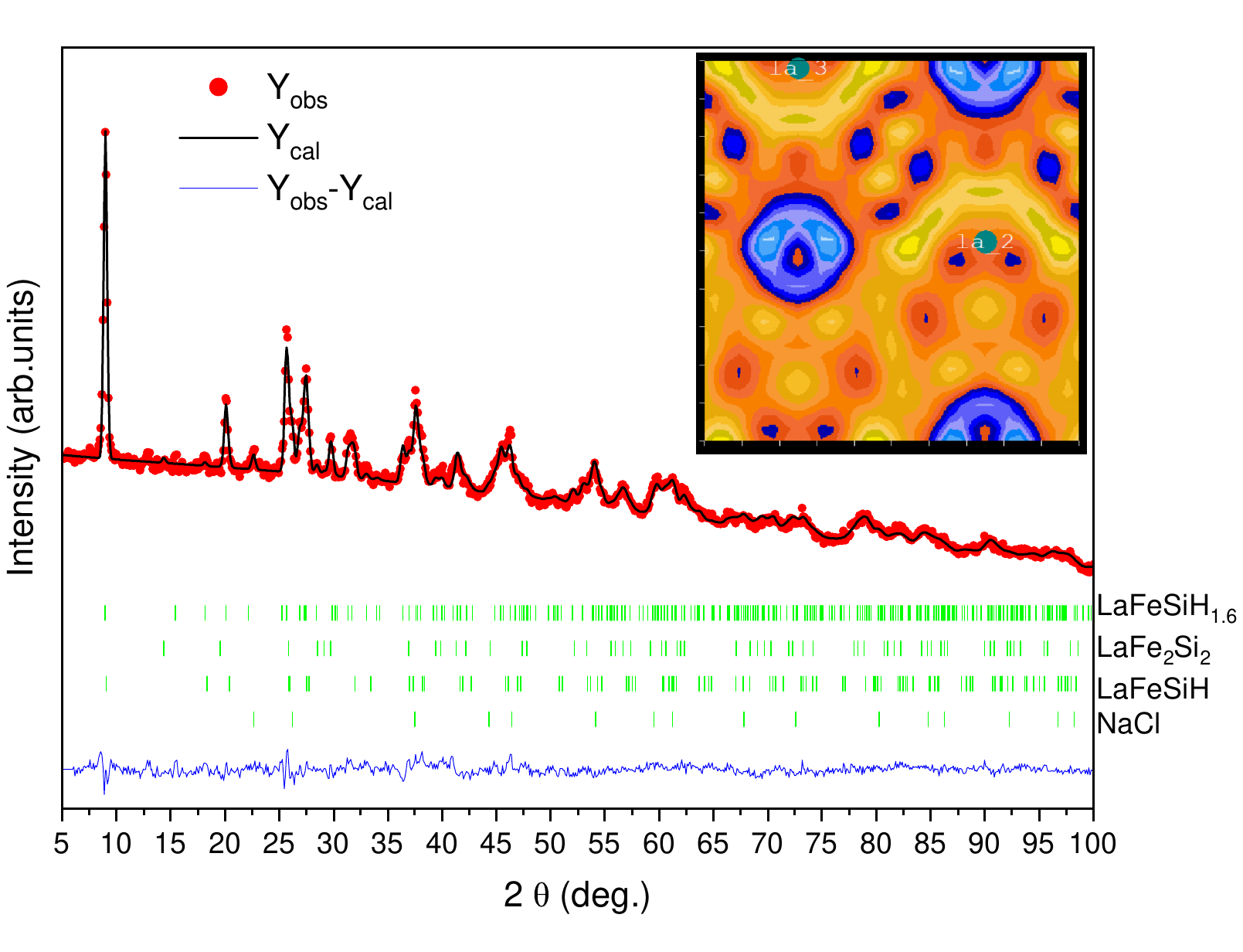}
    \caption{Rietveld fit of the neutron powder diffraction pattern collected on the LaFeSiH$_{1.6}$ including hydrogen position H2 localized in the Lanthanum plane (see table \ref{table:Pmba_ED_plus_NPD} and Fig.~\ref{fig:ED_Structures}). Impurity and secondary phases are also fitted in the model. Inset: Fourier difference map in the La plane (z~$\sim$~0.18 and $\sim$~0.82) calculated from NPD refinement using the first LaFeSiH structural model obtained from 3D ED as input, i.e without H2 site: negative nuclear density shown in blue evidences the H2 sites.}
    \label{fig:NPD}
\end{figure}

Building on the qualitative information obtained from electron diffraction patterns along zone axes, the structure of LaFeSiH$_{1.6}$ was further investigated using precession-assisted 3D ED. This approach aimed to achieve precise structural analysis, including the localization of hydrogen positions \cite{2017_Palatinus}. Using the Charge-Flipping algorithm, the structure solution proceeds without imposing symmetry, allowing the resulting density to be analyzed for symmetry. The analysis indicates the \textit{Pmab} space group as the most probable, despite a slightly less favorable agreement factor for the b-glide plane perpendicular to c. Assuming the \textit{Pmab} space group, the structure solution step allows for the determination of the atomic positions of the La, Fe and Si atoms. Subsequent structure refinement and the analysis of residual electron densities in the Fourier difference map provided some insights into the hydrogen distribution. A first hydrogen site (H1) can be located at the center of the (distorted) La$_{4}$ tetrahedra, consistent with the position observed in t-LaFeSiH \cite{2018_Bernardini,Layek_2024}, but featuring now different La-H bond lengths, as a consequence of the a- and b-axes differentiation. This structural model, including the H1 site with full occupancy, can be refined without any constraints leading to a global R(obs) value of about 12\% (Table \ref{table:Pmba_ED_plus_NPD}, left column). At this stage of the structural analysis, the observed lowering of symmetry and changes in lattice parameters, as revealed by both powder XRD and electron diffraction, suggest only a subtle deformation of the tetragonal LaFeSiH structure. However, the localization of the second hydrogen site (with a supposed occupancy of 0.6) cannot be conclusively determined based on the available 3D ED data or the examination of the Fourier difference maps.

To go further, we performed neutron powder diffraction to locate the last hydrogen position on a dedicated sample (containing around 70~wt~\% of o-LaFeSiH$_{1+x}$ as the main phase). Starting from the model obtained from ED, we examined the Fourier difference maps and were able to locate a position of negative nuclear density as seen in the inset of figure \ref{fig:NPD}. This is a tell-tale sign of a site occupied by hydrogen which has a negative scattering length. The position (confirmed by Rietveld refinement, see below) of this hydrogen (H2) is found localized in the lanthanum planes, almost directly above (or below) the Si along the c-axis, off-centered in a deformed square formed by neighboring La atoms, i.e. in a distorted \ch{La5Si} square bi-pyramid. This position of H2 away from the high symmetry position is realized through its systematic relative shift towards one of La site (in the a-b plane) along the b-axis, inducing its elongation (relatively to a-axis), and therefore explaining the observed orthorhombicity.

\begin{figure}[!htbp]
    \centering
    \includegraphics[width=0.8\linewidth]{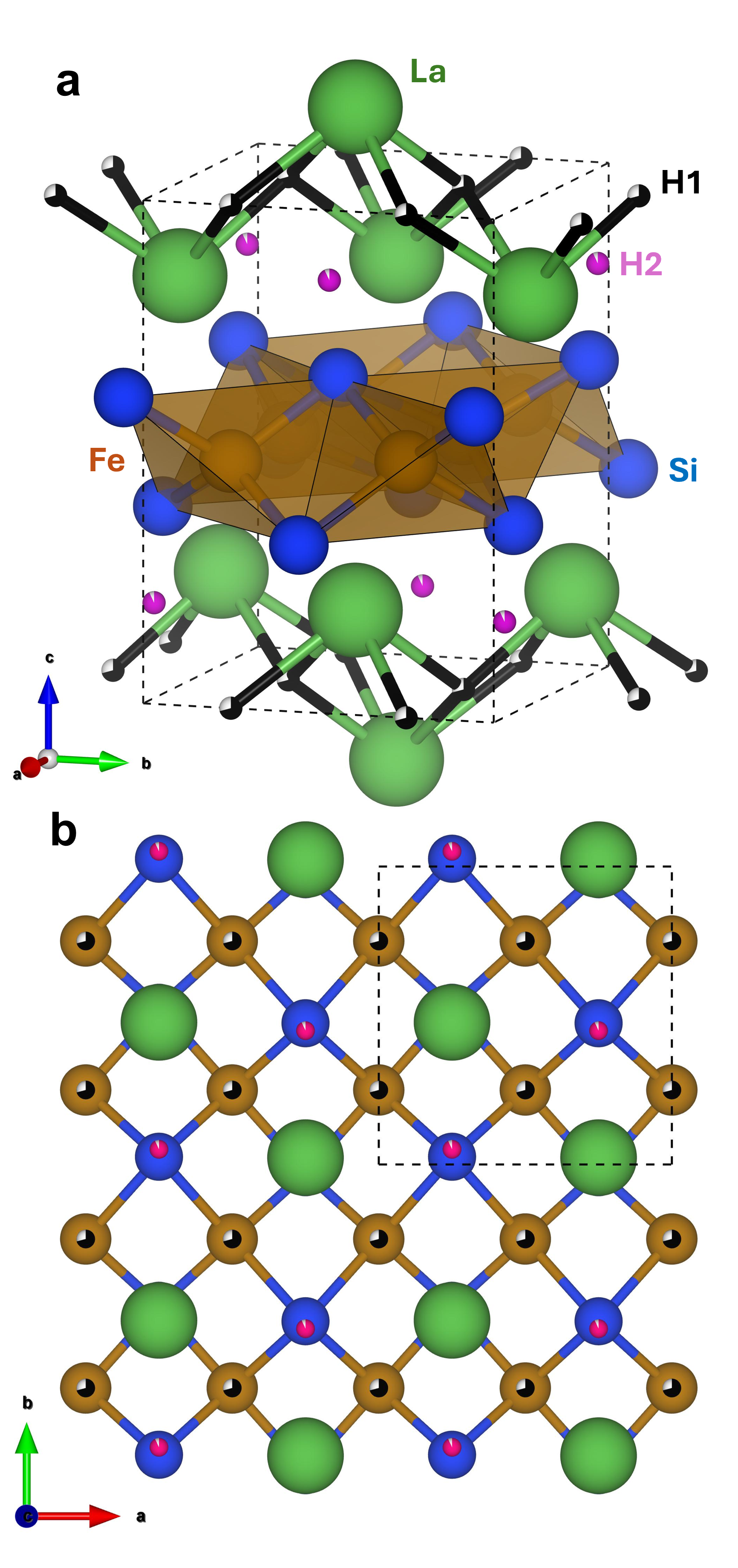}
    \caption{\textbf{(a)} Representation, using VESTA \cite{
    2011_Momma}, of the LaFeSiH$_{1.6}$ structural model obtained from neutron powder diffraction Rietveld refinement conducted in the \textit{Pmab} space group (see table~\ref{table:Pmba_ED_plus_NPD}). \textbf{(b)} A projection in the a,b-plane.
    }
    \label{fig:ED_Structures}
\end{figure}

We fitted the NPD pattern with a Rietveld model, as shown in figure \ref{fig:NPD}. Despite the high incoherent background coming from the presence of hydrogen in the sample, along with the somewhat low crystallinity of the samples, resulting from the high pressure intercalation synthesis route, we were able to fit the pattern in its full $2\theta$ range.
The scaling factors and lattice parameters of the main o-LaFeSiH$_{1+x}$ phase, secondary phases t-LaFeSiH and LaFe$_{2}$Si$_{2}$, and a NaCl impurity (from crucible) were refined. In addition, the atomic positions, isotropic atomic displacement parameters (ADP) and H1/H2 site occupancies (n) were refined for the main phase. Refining the ADP of hydrogen, or fixing it, does not strongly affect the relative positions and H1/H2 occupancies. The refined lattice parameters (a~=~5.753(1)~\AA, b~=~5.854(1)~\AA, c~=~8.083(2)~\AA) are in good agreement with the ones for the samples measured by XRD  (a~=~5.767(1)~\AA, b~=~5.853(1)~\AA, c~=~8.086(2)~\AA, Fig.~\ref{fig:XRD_synthesis}.b) or ED (table~\ref{table:Pmba_ED_plus_NPD}). Interestingly, the H1 site is found deficient with n(H1)~$\approx$~0.7(1). It was treated as fully occupied in the ED dynamical refinement, although it shows a relatively large isotropic ADP consistent with partial hydrogen deficiency. In contrast, the H2 site is nearly fully occupied with n(H2)~$\approx$~0.9(1), resulting in a composition, LaFeSiH$_{1.6}$, in agreement with the TGA results. We tested other models which could explain the orthorhombic distortion, for example placing the extra hydrogen (H2) in the Fe layer, in-between consecutive Fe atoms (along b-axis), however, none of them give a good fit of the NPD pattern.

The final crystal structure obtained, based on the NPD data refinement, is shown in figure \ref{fig:ED_Structures}.
By examining the local environment of the extra hydrogen site, we find that the nearest neighbor is the Si, with a distance of 1.57(4) \AA{}, whereas the La and H1 sites are located at average distances of 2.88(6) \AA{} and 2.46(5) \AA{} respectively. The Si-H distances are well within what has been observed in similar compounds where bonding distances range from 1.51-1.78 \AA, bounded by Ca$_{3}$Si$_{3}$H$_{3.4}$ and La(Fe$_{11.3}$Si$_{1.7}$)H$_{1.5}$~\cite{Auer_Ca3S3iH3.4,2000_Chacon,ROSCA201050}.
The presence of an additional H2 site is qualitatively consistent with the two components observed in the $^{29}$Si NMR spectrum. Its structural location, relatively close to both Fe and Si, could account for the larger shift of the additional spectral component. Moreover, the partial occupancies of the H1 and H2 sites introduce local disorder, which likely contributes to the broader lines observed in LaFeSiH$_{1.6}$ compared to LaFeSiH (Fig.~\ref{fig:NMR1}). If the H1 occupancy in the desorbed sample t-LaFeSiH$_{1+\delta}$ remains below 100\% as in o-LaFeSiH$_{1.6}$, it could explain the NMR differences relative to pristine LaFeSiH.

To put our findings into perspective, more than 150 compounds are known to crystallize in the ZrCuSiAs-type structure \cite{PottgenJohrendt}. Although these consist primarily of oxides and fluorides, a number of hydrides have been identified \cite{Chevalier2003,CHEVALIER2005529,PhysRevB.70.174408,PhysRevB.77.014414,MAHON2019701}. Several of these compounds can absorb more than 1 hydrogen pr. formula unit of the intermetallic, as we observe here for o-LaFeSiH$_{1.6}$. Notably, in CeNiSnH$_{1.8(2)}$, the extra hydrogen leads to a structural distortion from 	
orthorhombic CeNiSnH to a hexagonal system \cite{Chevalier2003}. This compound also shows a decomposition to CeNiSnH at moderate temperatures, 250~\textdegree{}C. 
Interestingly the LaFeSi(H/O/F) system is the IBS exhibiting the greatest chemical flexibility, and the discovery of o-LaFeSiH$_{1.6}$ further enforces that. This raises the question of whether hydrogen can be incorporated at sites other than the \textit{2b} (in the tetragonal lattice), in other iron-based superconductors and how such perturbation of structure and doping level might influence their physical properties.

\section{Conclusions}
We have synthesized a previously unreported hydride o-LaFeSiH$_{1.6}$ by hydrogenation of intermetallic LaFeSi, using thermal decomposition under high pressure of high-hydrogen-density materials as a hydrogen source. This layered compound is related to the superconducting LaFeSiH, but features an orthorhombic distortion induced by the additional hydrogen inserted into the layers. The resulting hydride, o-LaFeSiH$_{1.6}$, exhibits a semiconductor-like behavior and decomposes to t-LaFeSiH$_{1+\delta}$ upon mild heat treatment, recovering superconducting properties, however at a slightly different doping level from samples of LaFeSiH synthesized by topotactical intercalation methods. This is likely due to slight differences in hydrogen contents on the \textit{2b} H1 site (and residual H2 site).
Our work provides a method to reach the over-hydrogenated o-LaFeSiH$_{1.6}$ compound, which could be applicable to other materials, and importantly, demonstrates the chemical flexibility of the LaFeSiX [X~=~H, O, F] system. The discovery of o-LaFeSiH$_{1.6}$ represents an unprecedented case of tunability in the iron-based superconductors, with synthesis temperature being a key parameter, opening the possibility of exploring doping values beyond the limits ($x \simeq 0.8$) of systems like SmFeAsO$_{1-x}$H$_{x}$ \cite{2020Hosono}. As such, heavy hydrogenation may play a key role in uncovering new physics and in shedding light on the origin of superconductivity in the FeSi-layered family.

\section{Acknowledgement}
The authors thank Murielle Legendre (Institut N\'eel) for the machining of NaCl crucibles and magnetometry platform (Institut N\'eel) for their support during the measurements with the SQUID MPMS XL magnetometer. Work in Grenoble was supported by the Laboratoire d’excellence LANEF in Grenoble (ANR-10-LABX-51-01). 
This work was supported by the ANR-18-CE30-0018-03 Ironman grant.
We acknowledge the Institute Laue Langevin (ILL) for providing neutron beamtime at the D1B instrument.

\nocite{1992_WELTER,1998_WELTER,PHEJAR201695}

\bibliography{bib}

\clearpage

\section*{Supplemental information for: Stabilization of a non-superconducting, orthorhombic phase by over-hydrogenating
LaFeSiH}

\subsection{Superconductivity in t-LaFeSiH Anthracene based samples}
Tetragonal samples synthesized using Anthracene as a hydrogen donor were screened for superconductivity using magnetization and resistivity.  

\begin{figure}[h!]
    \centering
    \includegraphics[width=\linewidth,trim={1.9cm 1cm 3cm 2cm},clip]{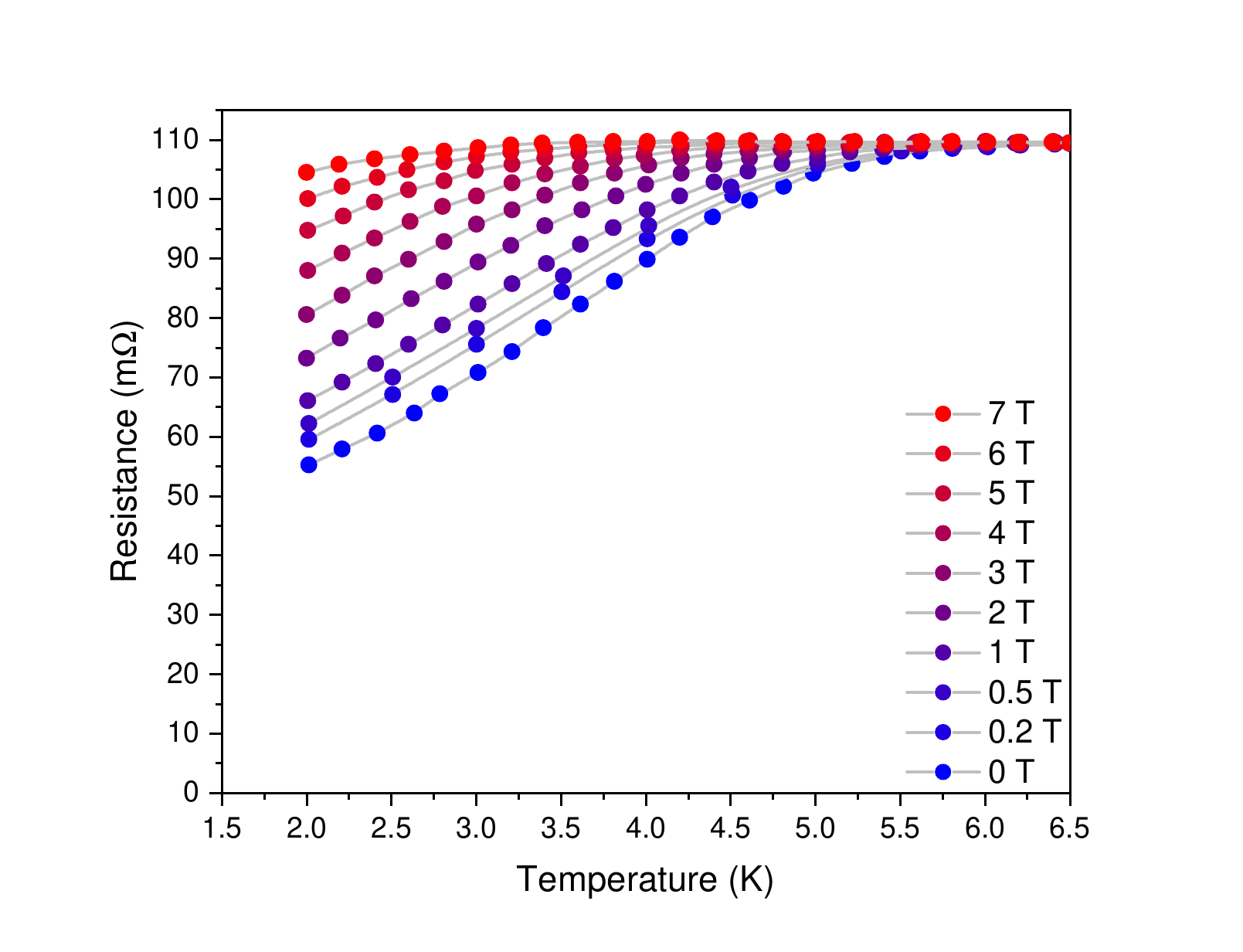}
    \caption{Resistivity data (in the low T range), under magnetic field (up to 7~T), recorded for a sample of t-LaFeSiH, synthesized at HP-HT, using Anthracene as an hydrogen source.}
    \label{fig:appendix:Antracene_resistivty}
\end{figure}

The resistivity data of typical t-LaFeSiH samples, shown in figure \ref{fig:appendix:Antracene_resistivty} presents a partial transition, with a drop of ~50~\% around 5.5~K. The origin of the lowering of T$_{c}$ is not clear, however it is likely that it is due to disorder resulting from the synthesis conditions, or slight changes in the hydrogen content. As shown on the Fig. the shift of T$_{c}$ with magnetic field, reaching T$_{c}$~$\approx$~3~K at 7~T is consistent with a superconducting transition.

\begin{figure}[h!]
    \centering
    \includegraphics[width=\linewidth,trim={1.9cm 1cm 2cm 1.5cm},clip]{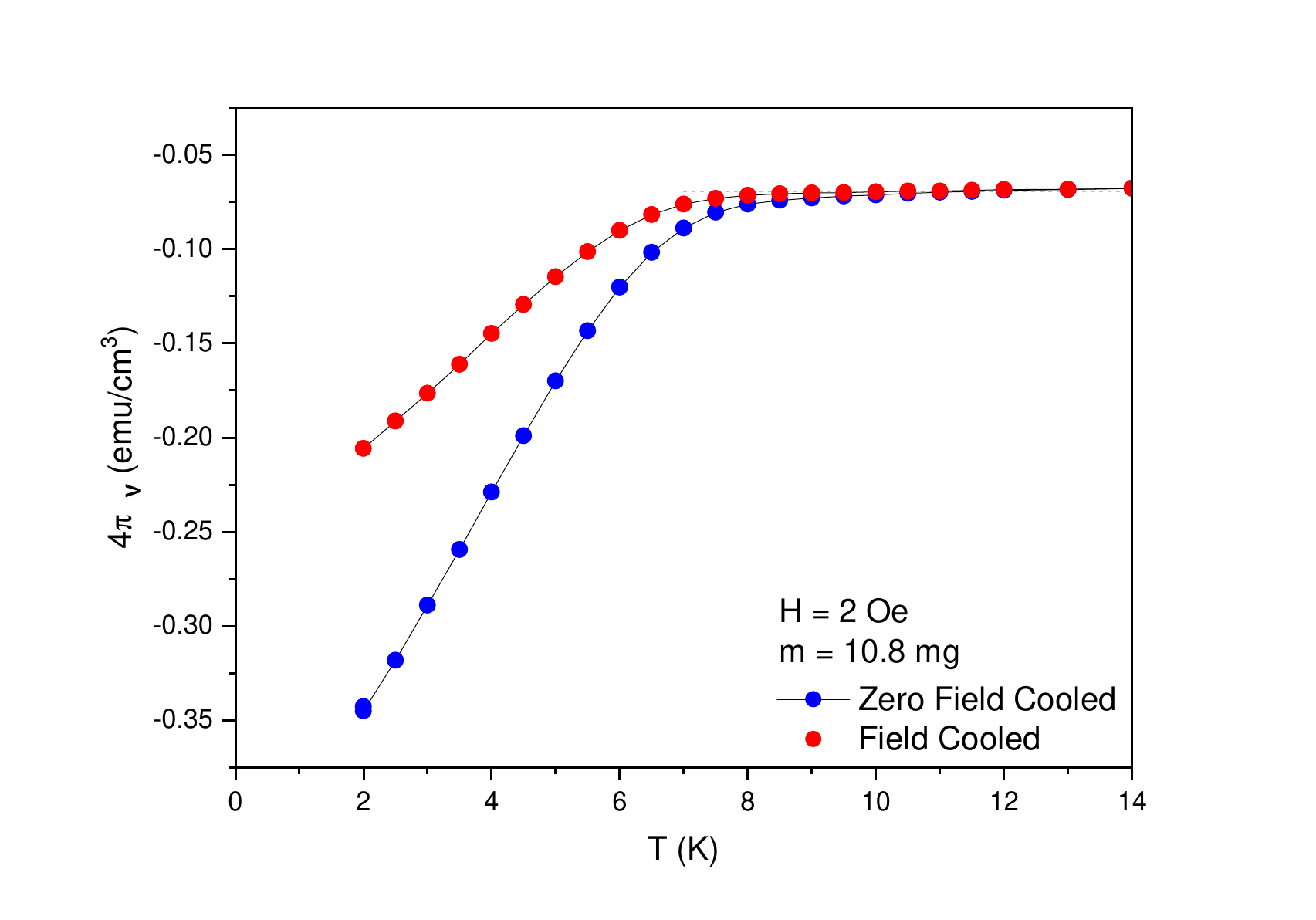}
    \caption{Magnetization data (in the low T range) measured at very low field (2~Oe) for a sample of t-LaFeSiH, synthesized at HP-HT, using Anthracene as an hydrogen source.}
    \label{fig:appendix:Antracene_magnetization}
\end{figure}

The corresponding magnetization data (measured at 2~Oe in a Metronique Ingenierie SQUID magnetometer) of such sample is shown in figure \ref{fig:appendix:Antracene_magnetization}. The transition has a shielding fraction estimated to be around 30~\% and has an onset around 6 K, consistent with the transition observed in resistivity. This shows that the transition is indeed bulk, however the incomplete resistive transition is likely due to grain boundary effects. 

\subsection{Superconductivity in desorbed t-LaFeSiH$_{1+\delta}$}

The samples recovered from dehydrogenation of o-LaFaSiH$_{1.5}$ to t-LaFeSiH$_{1+\delta}$ has a superconducting transition around T$_{c}$~=~8 K, as shown in its resistivity curve in the main text. 
Complementary magnetization measurements were performed on these samples, at very low magnetic field using the standard extraction method in a Metronique Ingenierie SQUID magnetometer and at larger field, up to 5~T, in a Quantum Design SQUID MPMS-XL magnetometer.

\begin{figure}[h!]
    \centering
    \includegraphics[width=0.8\linewidth]{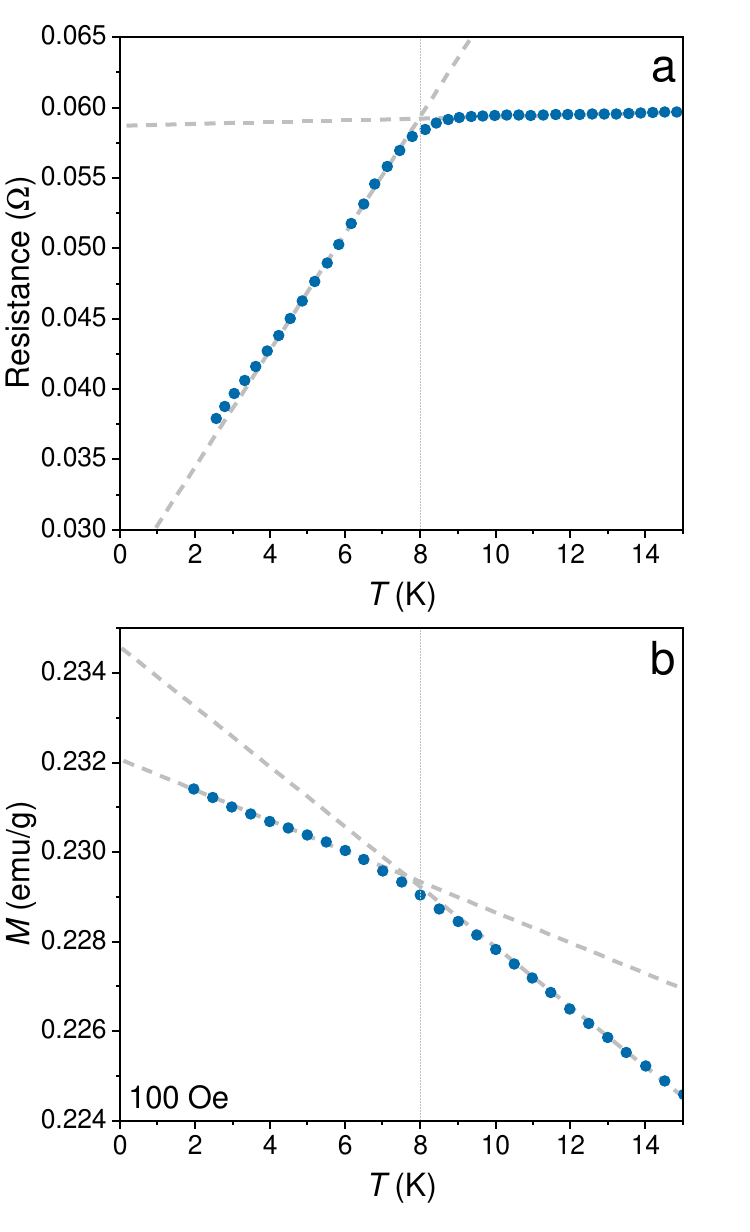}
    \caption{\textbf{a.} Resistivity data collected for the recovered t-LaFeSiH$_{1+\delta}$. \textbf{b.} Magnetization data measured at 100~Oe for a recovered t-LaFeSiH$_{1+\delta}$ sample.}
    \label{fig:SI:desorp_RandM}
\end{figure}

\begin{figure}[h!]
    \centering
    \includegraphics[width=\linewidth,trim={1.7cm 1.2cm 3cm 2cm},clip]{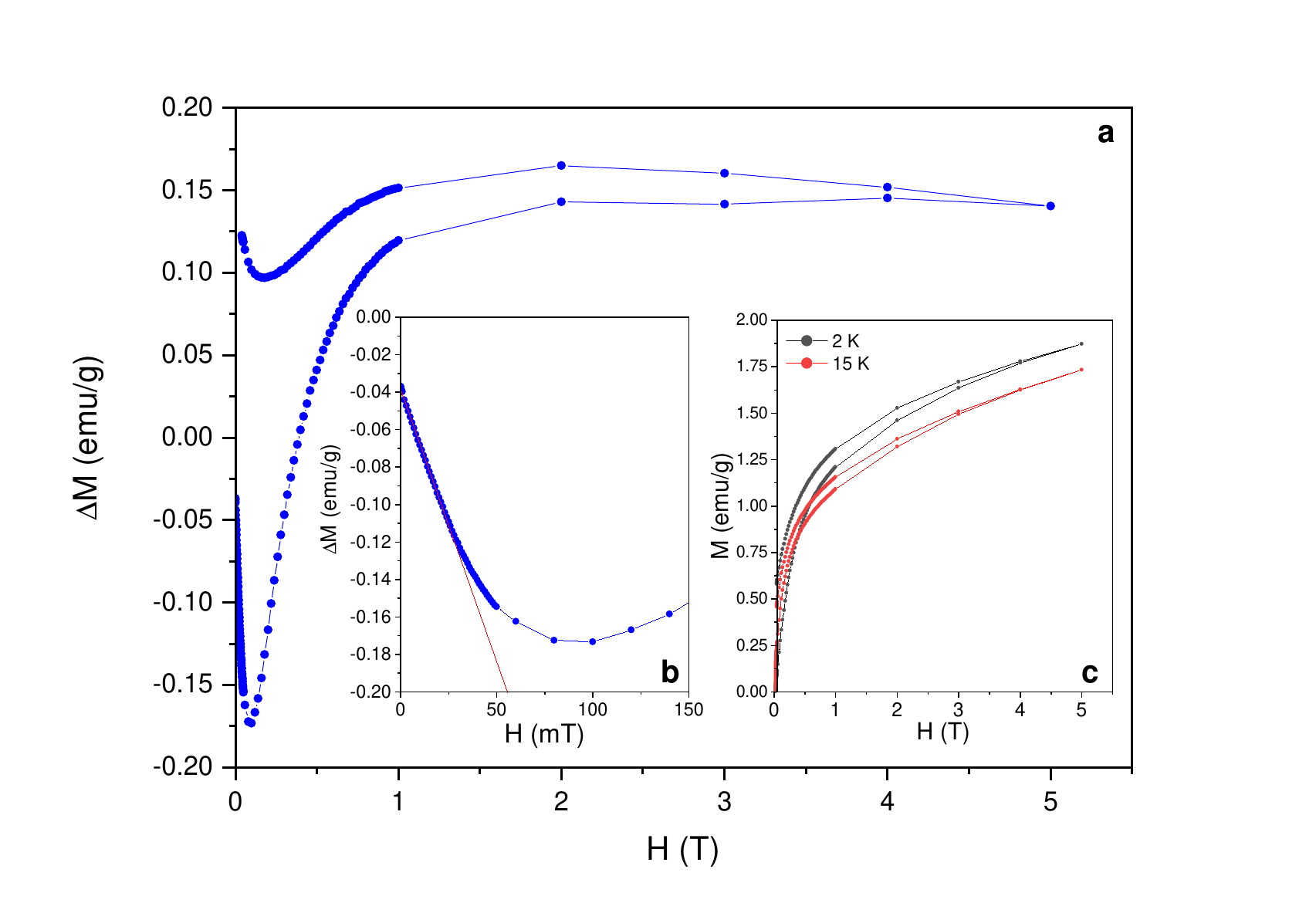}
    \caption{\textbf{a.} Magnetization data measured for a recovered t-LaFeSiH$_{1+\delta}$ sample. \textbf{b.} A zoom of the low field region shown in a.. \textbf{c.} The raw data from which the difference is obtained.}
    \label{fig:appendix:recovered_magnetization}
\end{figure}

Actually the temperature dependence of magnetization shows only a very weak kink in the data at the transition at T$_{c}$~=~8 K, as the magnetic signal is masked by the signal of ferromagnetic impurity, La(Fe$_{1-x}$Si$_{x}$)$_{13}$, as described in the paper proper. In figure \ref{fig:SI:desorp_RandM} the comparison of electrical resistance and magnetization measurements shows, in fact, that the transition observed in magnetization coincides with the one observed electrical resistance.
To verify that the transition observed in the magnetization measurement is indeed consistent with a superconducting transition, we measure M(H) up to 5 T at 2K and 15 K, i.e. above and below the transition. In figure \ref{fig:appendix:recovered_magnetization} we plot the magnetization data measured for a recovered sample, measured as a function of field. Figure \ref{fig:appendix:recovered_magnetization}.a shows the difference between the magnetization measured at 2~K and 15~K. We observe a clear diamagnetic response with a typical superconducting loop, which confirms the superconducting nature of the transition observed in resistivity. In figure \ref{fig:appendix:recovered_magnetization}.b we show a zoom of the difference signal, which clearly shows the linear diamagnetic behavior in the low field region.
In figure \ref{fig:appendix:recovered_magnetization}.c, the raw data (at 2~K and 15~K) which we took the difference of, to obtain the data shown in figure \ref{fig:appendix:recovered_magnetization}.a are plotted. We again note the significant magnetic background, originating mainly from La(Fe$_{1-x}$Si$_{x}$)$_{13}$H$_{y}$~\cite{PHEJAR201695}, masking the the superconducting contribution.

\medskip

\subsection{NMR peak fitting parameters}

 Fitting the NMR spectra of LaFeSiH$_{1.6}$, we found two solutions with opposite area ratios for the two peaks but the same goodness ($\chi^2 \approx 0.99$), depending on the initialization of parameters, as shown in Fig. 5. 

\begin{figure*}[!h]
    \centering
    \includegraphics[width=\linewidth]{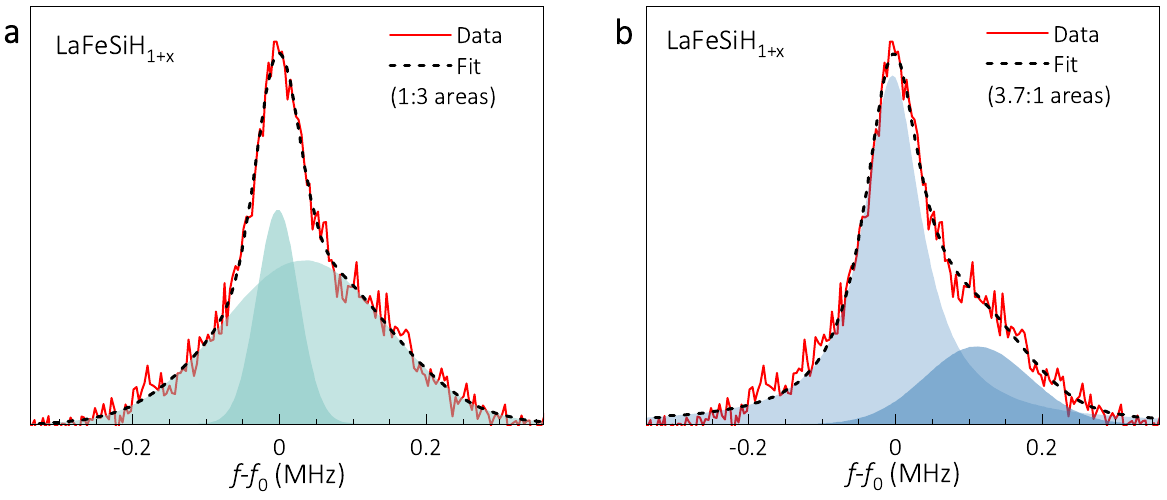}
    \caption{The $^{29}$Si NMR spectrum -- same as in Fig. 3 in the main text -- fit to two peaks of 1:3 (a) and 3.7:1 (b) relative areas. The fits in (a) and (b) are both obtained with unconstrained parameters, including areas, but they differ in the initialization of the parameters. In both cases, the given error bars are much smaller than the difference between (a) and (b), showing that the two-peak fit is practically non-identifiable due to strong parameter correlations.}
    \label{fig:NMR_sup}
\end{figure*}

\subsection{Magnetic field effects on electrical resistance}
\begin{figure*}[!h]
    \centering
    \includegraphics[width = \linewidth]{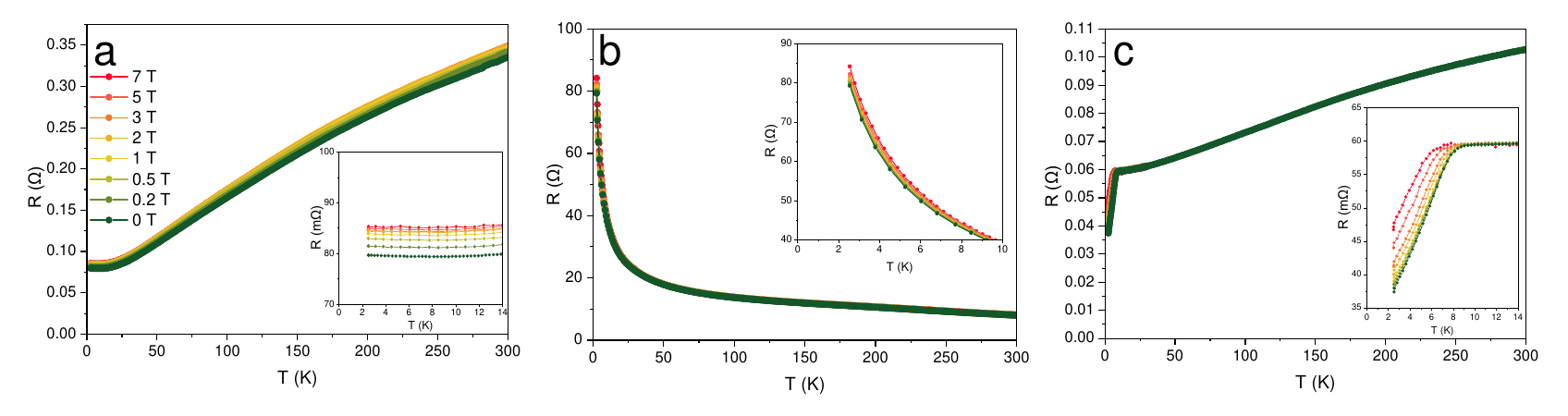}
    \caption{The electrical resistance of LaFeSi and derived compounds as a function of temperature and measured at different magnetic fields. (a): The measured resistance of LaFeSi. The absence of strong magneto-resistive effects is expected as LaFeSi is known to be an intermetallic Pauli paramagnet~\cite{1992_WELTER,1998_WELTER}. (b): The measured resistance of o-LaFeSiH$_{1.6}$. (c): The measured resistance of desorbed t-LaFeSiH$_{1+\delta}$. Upon increasing the field, the low-temperature drop shifts towards low temperature, as expected for a superconducting transition.}
    \label{fig:SI:fig:R(T)}
\end{figure*}

We measured the field dependence of the electrical resistance of the precursor LaFeSi, over-hydrogenated o-LaFeSiH$_{1.6}$, and superconducting, desorbed t-LaFeSiH$_{1+\delta}$. The resistance of these is plotted in figure \ref{fig:SI:fig:R(T)}. We observe no strong field dependence in LaFeSi nor in o-LaFeSiH$_{1.6}$, however, the field dependence of the transition seen in desorbed t-LaFeSiH$_{1+\delta}$ is consistent with a superconducting transtion.

\end{document}